\documentclass{elsart}

\usepackage{amssymb}

\begin{document}

\begin{frontmatter}



\title{Connection between the Green functions\\ of
 the supersymmetric pair  of Dirac Hamiltonians}


\author{Ekaterina Pozdeeva}

\address{Department of Quantum Field Theory,
  Tomsk State University, 36 Lenin Avenue, Tomsk, 634050, Russia}

\begin{abstract}
The Sukumar theorem about the connection between the Green functions
of the supersymmetric pair of the Schr\"o\-din\-ger Hamiltonians is
generalized to the case of the supersymmetric pair of the Dirac
Hamiltonians.
\end{abstract}

\footnotetext[0]{\textit{Email adress:} ekatpozdeeva@mail.ru
  (E. Pozdeeva).}
\renewcommand{\thempfootnote}{\arabic{mpfootnote}}
\footnotetext[1]{supported in part by the "Dynasty" Fund  and Moscow
International Center of Fundamental Physics.}
\end{frontmatter}

\section{Introduction}
In resent times a growing interest to applications of supersymmetric
quantum  mechanics \cite{Witten} in different fields of theoretical
and mathematical physics is noticed
\cite{Song2,Sukumar1985,Gomez,SUZKO,Pupasov}. Recently a special
issue of  J. Phys. A {\bf34}  was devoted to research work in
supersymmetric quantum mechanics.

It is well known that supersymmetric quantum mechanics is basically
equivalent to the Darboux \cite{Darbu} transformation and the
factorization properties of the Schr\"o\-din\-ger equation
\cite{Sukumar1985,Gomez,BagrovSamsonov}.  The Darboux transformation
method for the one-di\-me\-n\-sio\-nal sta\-tio\-na\-ry Dirac
equation is equivalent to the underlying quadratic supersymmetry and
the factorization properties of the Dirac equation
\cite{annphys2003v305p151,Deber}.

Though this method is widely used for the Schr\"o\-din\-ger equation
(see e. g. \cite{Sukumar,Samsonov}), its application to the Dirac
equation is studied much less \cite{Eurjphys,BAGROV,Pozdeeva}.

In the present paper we generalize   the Sukumar theorem
\cite{Sukumar} about the connection between the Green functions of
supersymmetric pair of the Schr\"o\-din\-ger Hamiltonians    to the
case of the Dirac Hamiltonians.

C.V. Sukumar have proved in \cite{Sukumar} the following relation:
\begin{eqnarray}
\label{1}
  \int_a^b[G_1(x,x,E)-G_0(x,x,E)]dx&=&\frac{1}{E-E_0},
\end{eqnarray}
where $G_1$, $G_0$ are the Green functions of supersymmetric  pair
of the Schr\"o\-din\-ger Hamiltonians
\begin{eqnarray}
  H_0&=&-\frac{d^2}{dx^2}+V_0(x),\qquad  H_1=LH_0L^{-1}, \qquad  L=\frac{d}{dx}-\frac{f'}{f},\\
  f&=&\psi_0(x), \qquad H_0\psi_0=E\psi_0,
\end{eqnarray}
with the  boundary conditions $\psi_n(a)=\psi_n(b)=0$
\begin{eqnarray}
H_0\psi_n&=&E_n\psi_n.
\end{eqnarray}

Above $\psi_0$ is the eigenfunction of initial ($H_0$) problem
without nodes.

We would like to stress that the construction of the Darboux
transformed
 of the  Schr\"o\-din\-ger Hamiltonian needs only one transformation
 function $f$  such that $H_0f=Ef$. The construction of the Darboux
 transformation of the Dirac problem needs two spinor functions
 $f_1$, $f_2$ such that
 \begin{eqnarray}
   H_0f_1&=&\lambda_1f_1,\qquad H_0f_2=\lambda_2f_2,\qquad
   \lambda_1\neq\lambda_2.
 \end{eqnarray}

Thus, four scalar functions are involved in the problem that made
this problem much more complicate. Another complication arises from
the fact that ``potential'' of the Dirac problem is a matrix
function. It is unclear is it possible to establish some relation
between the Green functions of the initial and the transformed
Hamiltonians. But for the potential of especial matrix structure
that described below this is possible. Below we consider this case.

The structure of the present paper is the following. In Section 2 we
give a new derivation of the Sukumar theorem for the
Schr\"o\-din\-ger case, different from original. In Section  3 we
generalize this theorem for the Dirac problem with especial choice
of matrix structure of interaction Hamiltonian. In Section 4 we
discuss the obtained results.

\section{Sukumar theorem}
In this Section  we give  the new derivation of Sukumar theorem for
the Schr\"o\-din\-ger  case because applied in this derivation
technique is more easily transferred  to the Dirac case. The Sukumar
problem is formulated as follows.

Let we have some initial Hamiltonian
\begin{eqnarray}
H_0&=&-\frac{d^2}{dx^2}+V_0(x),
\end{eqnarray}
where $V_0(x)$ is such that the set  of eigenfunctions of the $H_0$
\begin{eqnarray}
H_0\psi_n&=&E_n\psi_n
\end{eqnarray}
contains subset of this functions such that $\psi_n(a)=\psi_n(b)=0,$
where $a$, $b$, $(a<b)$ are some points on the real axis.

This subset of eigenfunction forms a complete system in the subspace
$\{X\}$ of all functions $\phi(x)$ such that $\phi(a)=\phi(b)=0.$
Last means that all $\phi(x)$ from this subspace can be presented in
the form: \begin{eqnarray} \phi(x)=\sum_nc_n\psi_n(x).
         \end{eqnarray}

Let us introduce the pair of the  solutions of equation

\begin{eqnarray}
H_0\phi_{1,2}(E,x)&=&E\phi_{1,2}(E,x)
\end{eqnarray}
such that $\phi_1(E,b)=0$, $\phi_2(E,a)=0$. They does not belongs to
the subspace $\{X\}$. Nevertheless  the following construction
\begin{eqnarray}
\label{10}
G(x,y,E)&=&\frac{\phi_1(x,E)\phi_2(y,E)\Theta(x-y)+\phi_2(x,E)\phi_1(y,E)\Theta(y-x)}{W},\\
W&=&\phi_1(x,E)\phi'_2(x,E)-\phi_2(x,E)\phi'_1(x,E)=const(E)
\end{eqnarray}
obeys the inhomogeneous equation
\begin{eqnarray}
(H_0-E)G(x,y,E)&=&\delta(x-y)
\end{eqnarray}
and boundary conditions
\begin{eqnarray}
G(b,y,E)&=&0,\qquad G_0(x,a,E)=0, \qquad x,y\in (a,b).
\end{eqnarray}
Thus, (\ref{10}) represent the Green function of the Hamiltonian
$H_0$ in the subspace $\{X\}$. Let us perform the Darboux
transformation
\begin{eqnarray}
\label{tildapsi}\psi&\rightarrow&L\psi=\tilde{\psi}=\psi'-\frac{f'}{f}\psi,\\
 H_0&\rightarrow&H_1=LH_0L^{-1},\qquad  f=\psi_0
\end{eqnarray}
and construct the Green function of the transformed Hamiltonian
\begin{eqnarray}
\tilde{G}(x,y,E)&=&\frac{\tilde{\phi}_1(x,E)\tilde{\phi}_2(y,E)\Theta(x-y)+\tilde{\phi}_2(x,E)\tilde{\phi}_1(y,E)\Theta(y-x)}{W},\\
\tilde{W}&=&\tilde{\phi}_1(x,E)\tilde{\phi}'_2(x,E)-\tilde{\phi}_2(x,E)\tilde{\phi}'_1(x,E)=\tilde{const}(E).
\end{eqnarray}
It is simple to prove that $\tilde{W}=(E-\lambda)W.$

The Sukumar theorem states
\begin{eqnarray}
\int_a^b[\tilde{G}(x,x,E)-G(x,x,E)]dx&=&\frac{1}{E-\lambda}.
\end{eqnarray}
Let us prove this theorem  in the manner different from one applied
in the paper \cite{Sukumar}. It is evident that
\begin{eqnarray}
\tilde{G}(x,x,E)&=&\frac{\tilde{\phi}_1(x)\tilde{\phi}_2(x)}{(E-\lambda)W}.
\end{eqnarray}

Using the definition (\ref{tildapsi}) we can rewrite
\begin{eqnarray}
  \tilde{\phi}_1(x)\tilde{\phi}_2(x)&\equiv&F'_1(x)-\frac{\phi_1(x)}{f}F'_2(x),\\
 \label{F} F_1(x)&=&\frac{\phi_1(x)}{f(x)}F_2(x),\qquad
  F_2(x)=f(x)\phi'_2(x)-f'(x)\phi_2(x).
\end{eqnarray}
Taking into account that functions $f$ and $\phi_2$ are the
eigenfunctions of the initial Hamiltonian $H_0$ it is easy to prove
that
\begin{eqnarray}
-\frac{\phi_1}{f}F'_2
&=&(E-\lambda)\phi_1\phi_2=(E-\lambda)WG_0(x,x,E).
\end{eqnarray}
From this observation it is  followed that
\begin{eqnarray}
\tilde{G}(x,x,E)&=&G_0(x,x,E)+\frac{F'_1(x)}{(E-\lambda)W}.
\end{eqnarray}
Thus, the following expression is correct:
\begin{eqnarray}
  \int_a^b[\tilde{G}(x,x,E)-G(x,x,E)]dx&=&\frac{1}{(E-\lambda)W}[F_1(b)-F_1(a)].
\end{eqnarray}
By the definition (\ref{F})

\begin{eqnarray}
  F_1(x)&=&\frac{\phi_1(x)}{f(x)}F_2(x)=\phi_1(x)(\phi'_2(x)-\frac{f'(x)}{f(x)}\phi_2(x)\\
 &=&\phi_1(x)\tilde{\phi}_2(x)\equiv\phi_2(x)\tilde{\phi}_1(x)+W.
\end{eqnarray}

From this consideration and the boundary properties of the
$\tilde{\phi}_{1,2}(x)$ it is followed that
\begin{eqnarray}F_1(b)-F(a)=W\end{eqnarray} and
\begin{eqnarray}
\int_a^b[\tilde{G}(x,x,E)-G(x,x,E)]dx&=&\frac{1}{E-\lambda}\end{eqnarray}
what is just the Sukumar theorem.
\section{Dirac problem}
In some cases the one-dimensional four component Dirac equation
admit the two component representation
\begin{eqnarray}
  h_0\psi(x)&=&E\psi(x),\qquad \psi^T(x)=[\psi_1(x),\psi_2(x)],\\
  h_0&=&i\sigma_2\partial_x+\sigma_3(m+S(x))+\sigma_1U(x)+V(x)
\end{eqnarray}
where $\sigma_{1,2,3}$ are the Pauli matrices.

We don't discuss here the physical situation that can be described
by this equation. We would like only to note that even this
oversimplified problem is not simple for the consideration.

The one problem is the generalization of the one-component
Schr\"o\-din\-ger problem with the boundary conditions to the
two-component Dirac problem with the boundary conditions is
following:  we can put the conditions $\psi(a)=\psi(b)=0$ only  on
one component $\psi_{1,2}$ of spinor satisfying the Dirac equation.

The another problem is the following: the  spinor functions obtained
after the Darboux transformation  as a rule do not satisfy the same
boundary conditions as the initial spinors functions.

The full consideration  of  this problems will be present later.
Here we only would like  to note that the case $S(x)=V(x)=0$ admit
analytical solution problem of the connection  between Green
functions of the initial and the Darboux transformed problem
(supersymmetric partners) problem.

So we look like for the solution of the Dirac equation
\begin{eqnarray}
h_0\psi&=&E\psi,\quad \psi=(\psi_1,\psi_2)^T \label{equation}\\
h_0&=&i\sigma_2\partial_x+m\sigma_3+U\sigma_1
\end{eqnarray}
with boundary conditions. This conditions can be of two types
\begin{eqnarray}
(i)\quad \psi_1(a)&=&\psi_1(b)=0\label{cond1},\\
(ii)\quad \psi_2(a)&=&\psi_2(b)=0\label{cond2}.
\end{eqnarray}
Let us rewrite the Dirac equation in the component form:
\begin{eqnarray}
  \phi'_2+U\phi_2&=&(E-m)\phi_1 \label{conponentequation1}\\
  -\phi'_1+U\phi_1&=&(E+m)\phi_2\label{conponentequation2}.
\end{eqnarray}
It is essential for the following that  in the case (\ref{cond1})
among others solutions of the Dirac equation there exist especial
solution of the form:
\begin{eqnarray}
\phi_1(x)&\equiv&0,\qquad \phi_2(x)=exp(-\int^xU(x')dx'),\qquad E=-m
\end{eqnarray}
and in the case (\ref{cond2})
\begin{eqnarray}
\phi_2(x)&\equiv&0,\qquad \phi_1(x)=exp(\int^xU(x')dx'),\qquad E=m.
\end{eqnarray}
First of all, consider the Dirac equation with the boundary
conditions (\ref{cond1}). This condition chooses among all solutions
of eigenvalue problem (\ref{equation}) the subset of solutions with
discrete spectrum
\begin{eqnarray}
h_0\psi^{(n)}&=&E_n\psi^{(n)},\qquad
\psi_1^{(n)}(a)=\psi_1^{(n)}(b)=0.
\end{eqnarray}
The solution $h_0v=-mv$ belongs to this subset.

Let us call $u$ the solution of system of equations
(\ref{conponentequation1}) and (\ref{conponentequation2}) with the
such eigenvalue $E$ that is one of the neighboring on $E=-m.$ Thus,
we have two possibilities $E>-m$ and $E<-m.$ Denote the eigenvalue
of chosen solution by $\lambda$ $h_0u=\lambda u.$

 Let us construct the Darboux transformation of the form:
 \begin{eqnarray}
L&=&\partial_x-\hat{u}_xu^{-1},\qquad \hat{u}_x=\partial_x\hat{u},\qquad \hat{u}=\left(
\begin{array}{cc}
v_1 & u_1 \\
v_2 & u_2 \\
\end{array}
\right),\\
   v_1&\equiv&0,\qquad v_2=exp(-\int^xU(x')dx')
 \end{eqnarray}
and apply it to any solution of the equation (\ref{equation})
$\tilde{\psi}=L\psi$.

Then with the help of a simple algebra it is easy to obtain
\begin{eqnarray}
\label{tildapsi1}\tilde{\psi}_1&=&\partial_x\psi_1-\frac{f'}{f}\psi_1, \quad f=u_1, \\
\label{tildapsi2}\tilde{\psi}_2&=&(E-\lambda)\psi_1.
\end{eqnarray}
Let us denote $\psi^{(L)}$, $\psi^{(R)}$ the solution of the initial
equation (\ref{equation}) that obey the following boundary
conditions:
\begin{eqnarray}
\psi^{(L)}(a)&=&0,\qquad  \psi^{(R)}(b)=0.
\end{eqnarray}
Then from (\ref{tildapsi1}), (\ref{tildapsi2}) it is easy obtained
that
\begin{eqnarray}
\tilde{\psi}^{(L)}(a)&=&0,\qquad  \tilde{\psi}^{(R)}(b)=0.
\end{eqnarray}
Let us construct the Green functions of the initial ($h_0$) and the
transformed Hamiltonians ($h_1=Lh_0L^{-1}$). These functions can be
expressed in the terms $\psi^{(L)}$, $\psi^{(R)}$ (for $h_0$) and
$\tilde{\psi}^{(L)}$, $\tilde{\psi}^{(R)}$ (for $h_1$) respectively
\begin{eqnarray}
G_0(x,y,E)&=&\frac{\psi^{(R)}(x)\psi^{(L)T}(y)\Theta(x-y)+\psi^{(L)}(x)\psi^{(R)T}(y)\Theta(y-x)}{W},\\
  W&=&\psi^{(R)}_1(x)\psi^{(L)}_2(x)-\psi^{(L)}_1(x)\psi^{(R)}_2(x)=const(E)
\end{eqnarray}
and similarly,
\begin{eqnarray}
G_1(x,y,E)&=&\frac{\tilde{\psi}^{(R)}(x)\tilde{\psi}^{(L)T}(y)\Theta(x-y)+\tilde{\psi}^{(L)}(x)\tilde{\psi}^{(R)T}(y)\Theta(y-x)}{\tilde{W}},\\
 \tilde{W}&=&\tilde{\psi}^{(R)}_1(x)\tilde{\psi}^{(L)}_2(x)-\tilde{\psi}^{(L)}_1(x)\tilde{\psi}^{(R)}_2(y)=(E-\lambda)(E+m)W.
\end{eqnarray}
Let us introduce the quantities
\begin{eqnarray}
\label{49}I_0(x,E)&=&trG_0(x,x,E)=\frac{\psi_1^{(L)}(x)\psi_1^{(R)}(x)+\psi_2^{(L)}(x)\psi_2^{(R)}(x)}{W} \\
\label{50}I_1(x,E)&=&trG_1(x,x,E)=\frac{\tilde{\psi}_1^{(L)}(x)\tilde{\psi}_1^{(R)}(x)+\tilde{\psi}_2^{(L)}(x)\tilde{\psi}_2^{(R)}(x)}{\tilde{W}}.
\end{eqnarray}

  Let us express the numerator of the $I_1(x,E)$
in terms of functions $\psi^{(L)}$, $\psi^{(R)}$
\begin{eqnarray}
&&\tilde{\psi}_1^{(L)}(x)\tilde{\psi}_1^{(R)}(x)+\tilde{\psi}_2^{(L)}(x)\tilde{\psi}_2^{(R)}(x)=A+B \\
&&A=(\psi_1^{'(L)}-\frac{f'}{f}\psi_1^{(L)})(\psi_1^{'(R)}-\frac{f'}{f}\psi_1^{(R)}) \\
&&B=(E-\lambda)^2\psi_1^{(L)}\psi_1^{(R)}.
\end{eqnarray}
Using the similar trick as above (see Section 2) one can obtain
\begin{eqnarray}
A&=&F'_1-\frac{\psi_1^{(L)}}{f}F'_2, \\
  F_1&=&\psi_1^{(L)}(\psi_1^{'(R)}-\frac{f'}{f}\psi_1^{(R)}),\\
  F_2&=&f\psi_1^{'(R)}-f'\psi_1^{(R)}).
\end{eqnarray}
 Using the Dirac equations in order to exclude derivatives
 $\psi^{('R)}$ and $f'$ from expression for $F_2$ and once more for
excluding derivatives from expression for $F'_2$ after simple but
rather cumbersome calculations one can obtain
\begin{eqnarray}
-\frac{\psi_1^{(L)}}{f}F'_2&=&(E^2-\lambda^2)\psi_1^{(L)}\psi_1^{(R)}.
\end{eqnarray}
As a results we have
\begin{eqnarray}
A+B&=&F'_1+2E(E-\lambda)\psi_1^{(L)}\psi_1^{(R)}.
\end{eqnarray}
Let us now  to consider the difference
\begin{eqnarray}
\Delta I&=&I_1-I_0=trG_1(x,x,E)-trG_0(x,x,E).
\end{eqnarray}
With usage of eq. (\ref{49}), (\ref{50}) we have
\begin{eqnarray}
\Delta
I&=&\frac{F'_1}{(E-\lambda)(E+m)W}+\frac{(E-m)\psi_1^{L}\psi_1^{(R)}-(E+m)\psi_2^{(L)}\psi_2^{(R)}}{(E+m)W}.
\end{eqnarray}
Again by the usage of the Dirac equation for $\psi^{(L,R)}$ it is
easily to obtain
\begin{eqnarray}
(E-m)\psi_1^{L}\psi_1^{R}-(E+m)\psi_2^{(L)}\psi_2^{(R)}&=&(\psi_1^{(L)}\psi_1^{(R)})'=F_3'.
\end{eqnarray}
Taking into account identity
\begin{eqnarray}
psi_1^{(L)}\psi_2^{(R)}=\psi_1^{(R)}\psi_2^{(L)}+W\end{eqnarray} and
boundary conditions
\begin{eqnarray}\psi_1^{(L)}(a)=\psi_1^{(R)}(b)=0\end{eqnarray} it is easy to
see that \begin{eqnarray} \label{60} \int_a^bF'_3dx&=&W.
\end{eqnarray}
Let us briefly discuss the quantity
\begin{eqnarray}
F_1(x)&=&\psi_1^{(L)}(\psi_1^{'(R)}-\frac{f'}{f}\psi_1^{(R)})\equiv\psi_1^{(L)}\tilde{\psi}_1^{(R)}.
\end{eqnarray}
It can be rewritten in the form:
\begin{eqnarray}
F_1(x)&=&\psi_1^{(L)}\psi_1^{'(R)}-\psi_1^{'(L)}\psi_1^{(R)}+\psi_1^{(R)}\tilde{\psi}_1^{(L)},
\end{eqnarray}
that after  excluding of derivatives from the last expression with
the help of the Dirac equation can be  presented in the form:
\begin{eqnarray}
F_1(x)&=&(E+m)W+\psi_1^{(R)}\tilde{\psi}_1^{(L)}.
\end{eqnarray}

 Taking into
account this observation it easily to obtain
\begin{eqnarray}\int_a^bF'_1=(E+m)W.\end{eqnarray}  Combining  this result with the eq. (\ref{60}), we
obtain the final result:
\begin{eqnarray}
\int_a^b[tr G_1(x,x,E)-tr G_0(x,x,E)]dx
&=&\frac{1}{E-\lambda}+\frac{1}{E+m}.
\end{eqnarray}

In the case (\ref{cond2}) the similar calculations lead to the
result:
\begin{eqnarray}
\int_a^b[tr
G_1(x,x,E)-trG_0(x,x,E)]dx&=&\frac{1}{E-\lambda}+\frac{1}{E-m}.
\end{eqnarray}
This is the generalization of the Sukumar theorem for the Dirac
case.
\section{Discussion}
It is well known that the Green function of any Hamiltonian can be
represented in the form:
\begin{eqnarray}
G(x,y,E)&=&\Sigma_n\frac{\psi_n(x)\psi_n^T(x)}{E_n-E},
\end{eqnarray}
\begin{eqnarray}
\int tr(\psi_n(x)\psi_n^T(x))dx&=&1
\end{eqnarray}
(the spectral representation). In the integral
\begin{eqnarray*}I=\int tr G(x,x,E)dx\neq\pm\infty\end{eqnarray*},
then
\begin{eqnarray*}I=\sum_n\frac{1}{E_n-E}\end{eqnarray*} and the Sukumar and our
results are trivial since they reflect well known observation that
Darboux  transformation delete from the spectrum of the initial
Hamiltonian the terms, corresponding to the states, that used for
the construction of the transformation matrix $u$.

In the case when \begin{eqnarray*}|I|=\pm\infty\end{eqnarray*} this
result is not trivial as it is seen from the results of the
calculation of
\begin{eqnarray*}
\int_{-\infty}^{\infty}(G_1(x,x,E)-G_0(x,x,E))dx
\end{eqnarray*}
for some especial choices of the transformation function $f$
(``Schr\"o\-din\-ger case'') or transformation matrix $\hat{u}$
(``Dirac case'').

 In the both considered cases the difference
\begin{eqnarray*}
\int_{-\infty}^{\infty}tr(G_1(x,x,E)-G_0(x,x,E))dx
\end{eqnarray*}
contains beside terms $(E-\lambda)^{-1}$ (``Schr\"o\-din\-ger
case'') or $(E-\lambda_1)^{-1}+(E-\lambda_2)^{-1}$ (``Dirac case'')
also the effects of ``nonequality'' of two initial normalization
constant of the initial and the transformed problem.
\section*{Acknowledgments}

The author is grateful to the Joint Institute for Nuclear Research
(Dubna, Moscow region)  for hospitality during this work. This work
was supported in part by the ``Dynasty'' Fund and Moscow
International Center of Fundamental Physics.

\end{document}